\documentclass[reprint, superscriptaddress,
 amsmath,amssymb,
 aps,
 pra,
]{revtex4-2}
\usepackage[usenames,dvipsnames]{color}

\usepackage{graphicx}
\usepackage{dcolumn}
\usepackage{bm}

\begin{document}

\preprint{APS/123-QED}

\title{Collinear Optical Two-Dimensional Coherent
Spectroscopy with Fluorescence Detection at 5 kHz
Repetition Rate}

\author{Stephen Revesz}
\affiliation{
 Department of Physics, Florida International University, Miami, FL 33199, USA
}
\author{Rustam Gatamov}
\affiliation{
 Department of Physics, Florida International University, Miami, FL 33199, USA
}
\author{Adolfo Misiara}
\affiliation{
Department of Physics, Florida International University, Miami, FL 33199, USA
}
\author{Hebin Li}
 \email{Corresponding author: hebinli@miami.edu}
\affiliation{
Department of Physics, Florida International University, Miami, FL 33199, USA
}
 \affiliation{Department of Physics, University of Miami, Coral Gables, FL 33124, USA}




\date{\today}

\begin{abstract}
Optical two-dimensional coherent spectroscopy (2DCS) is a powerful ultrafast spectroscopic technique that can greatly benefit from the unique features of a femtosecond laser operating at a kHz repetition rate. However, isolating specific nonlinear signal in the collinear geometry, especially with a kHz laser, presents challenges. We present an experimental implementation of optical 2DCS in a collinear geometry using a femtosecond laser operating at a 5 KHz repetition rate. The desired nonlinear signal is selectively extracted in the frequency domain by lock-in detection. Both pump-probe and optical 2DCS experiments were conducted on a rubidium vapor. The signal-to-noise ratio of pump-probe and 2DCS spectra was characterised under various experimental parameters. The study highlights the importance of the lock-in reference frequency in overcoming the limitations imposed by low repetition rates, thereby paving the way for the application of collinear optical 2DCS in material systems requiring kHz femtosecond lasers. 
\end{abstract}

\maketitle


Optical two-dimensional coherent ultrafast spectroscopy (2DCS) has proven to be a powerful technique for probing many-body interactions and dynamics in atomic ensembles \cite{Dai2012a, Gao:16, PhysRevLett.120.233401, Yu2018, Yu2019, Liang2021, Yan2022, PhysRevLett.128.103601, Landmesser23, Liang2022}, 2D materials \cite{Moody2015,Titze2018, PhysRevX.13.011029}, diamond color centers \cite{Smallwood2021}, and other systems \cite{MDCSBook2023}. Optical 2DCS can be implemented in either a noncollinear beam geometry \cite{Brixner:04, COWAN2004184, Bristow2009, Turner2011}, where the excitation pulses are in different directions, or a collinear geometry \cite{Tekavec2007, Nardin2013} with co-propagating excitation pulses. In a noncollinear geometry, such as the box geometry, the desired nonlinear signal is spatially separated from the excitation pulses and other nonlinear signals, resulting in a high signal-to-noise ratio (SNR). In a collinear geometry, the co-propagating pulses can be tightly focused by an objective lens to achieve a high spatial resolution. The excitation pulses are usually provided by a femtosecond (fs) laser. In some experiments, a Ti:Sapphire regenerative amplifier operating at a kHz (1-10 kHz) repetition rate might be preferred due to the need for high pulse energy, wavelength tunability, low average power to avoid heating effects, and long waiting time between shots for the sample to fully relax back to the ground state. Although many optical 2DCS implementations in the box geometry use a kHz laser system \cite{Brixner:04, COWAN2004184}, it remains challenging to perform collinear 2DCS at a kHz repetition rate.

In optical 2DCS experiments, the nonlinear signal from specific excitation pathways needs to be isolated. The signal is spatially separated in the box geometry according to the phase matching condition. However, the method does not work in the collinear geometry where all signals propagate in the same direction. The desired signal can be selected by using a multi-step phase cycling procedure \cite{Wagner:05} or frequency filtering \cite{Tekavec2007, Nardin2013}. In collinear pump-probe and 2DCS experiments, each excitation pulse is phase-modulated by an acousto-optic modulator (AOM) at a unique frequency where each pulse is effectively frequency tagged. The desired signal due to a specific excitation pulse sequence is phase modulated and can be isolated in the frequency domain by a lock-in amplifier at the proper reference frequency obtained by mixing the modulation frequencies of excitation pulses. This phase modulation technique typically requires a high lock-in reference frequency to minimize the effect of environmental noise and achieve a good SNR. However, the higher end of lock-in reference frequency is limited by the laser repetition rate ($\nu_{rep}$) to ensure there is at least one shot within each cycle of lock-in detection. Moreover, to avoid aliasing, the lock-in reference frequency ($\nu_{LI}$) is limited by the Nyquist theorem, which specifies: $\nu_{LI} < 0.5\nu_{rep}$. Fundamentally, the choice of lock-in reference frequency $\nu_{LI}$ is limited by the noise floor in the experiment and the laser repetition rate $\nu_{rep}$. It becomes experimentally challenging to satisfy this requirement and achieve a good SNR with the low repetition rate of a kHz laser. In a previous work by Bruder et al. \cite{Bruder:18}, ultrafast pump-probe spectroscopy measurements were conducted on a rubidium (Rb) vapor in a collinear geometry to study the effects of low lock-in reference frequencies and undersampling on the SNR. 

\begin{figure*}[thb]
    \centering
    \includegraphics[width=\textwidth]{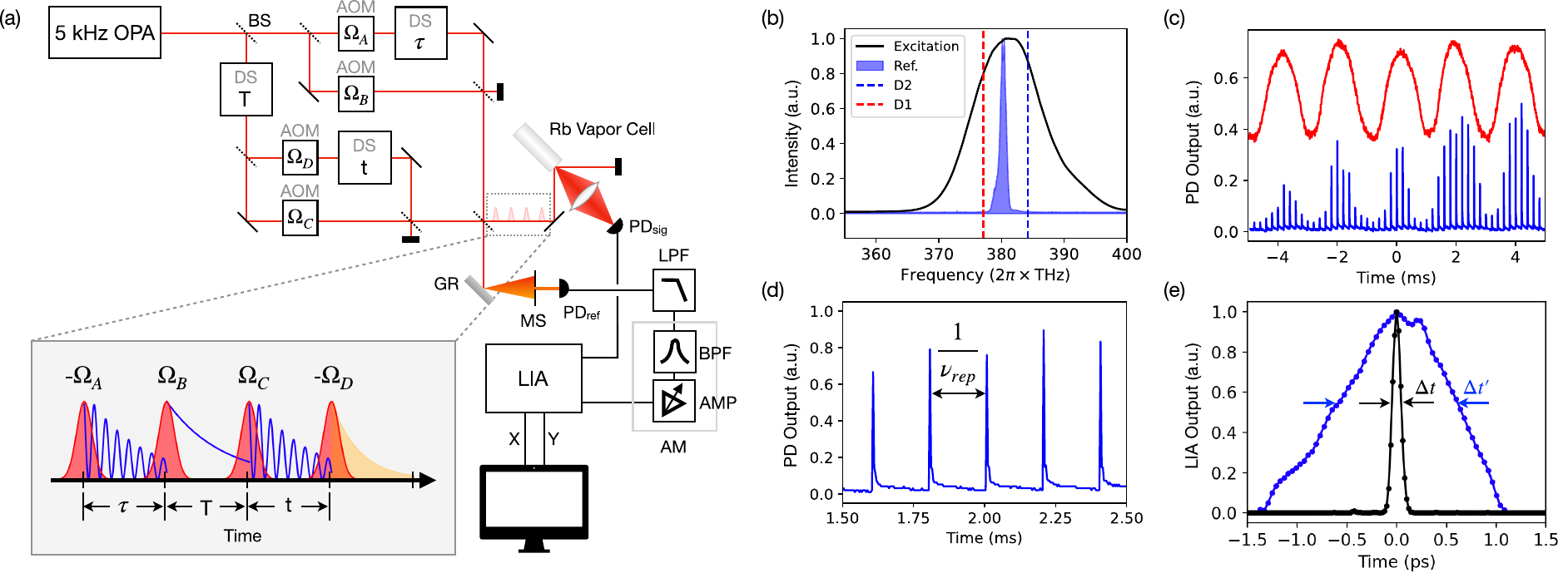}
    \caption{(a) Schematic of the collinear 2DCS spectrometer. BS: Beamsplitter. DS: Delay stage. AOM: Acousto-optic modulator. GR: Grating. PD: Photodetector. MS: Manual slit. LPF: Low-pass filter. AM: Audio mixer. BPF: Bandpass filter. AMP: Amplifier. LIA: Lock-in amplifier. The pulse ordering scheme utilized is shown in the inset. (b) The spectra of the OPA output (black) and the shaped pulse (blue-shaded area) measured with a spectrometer. The Rb $D_{1}$ ($D_{2}$) transition frequency is indicated by the red (blue) dashed line. (c) The output of PD$_{ref}$ showing the amplitude modulation due to beating between $A$ and $B$, before (blue) and after (red) passing through a 2-kHz LPF. The traces are offset for clarity. (d) Zoom in on the blue trace in (c) between 1.5 to 2.5 ms to show the $\nu_{rep} = 5$ kHz pulse train. (e) Auto-correlation spectra of the pulse pair $A$ and $B$, with (blue dots) and without (black dots) passing through the pulse shaper.} 
    \label{fig:1}
\end{figure*}

In this Letter, we report the experimental implementation of collinear 2DCS measurements on a Rb vapor with a $\nu_{rep}=5$ kHz fs laser. The nonlinear fluorescence signal is isolated by lock-in detection at the reference frequency $\nu_{LI} < 0.5\nu_{rep}$ to perform both pump-probe and 2DCS measurements. We characterize the SNR dependence on the lock-in reference frequency $\nu_{LI}$ for pump-probe and one-quantum rephasing (S1) 2DCS. The study details how to overcome the limitations imposed by low repetition rate fs laser for fluorescence detection collinear 2DCS.

\begin{figure}[!ht]
    \centering
    \includegraphics[width=\columnwidth]{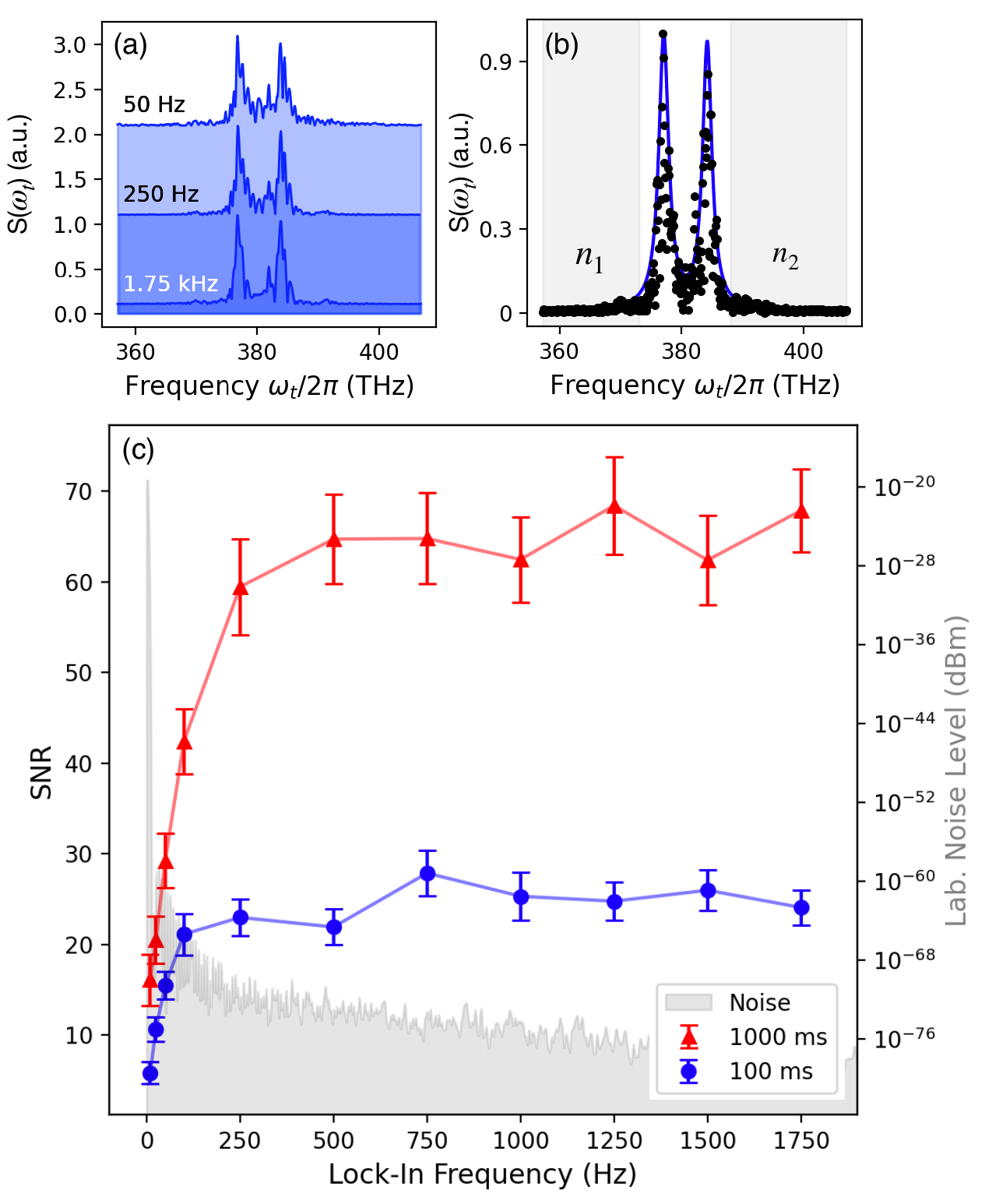}
    \caption{(a) Pump-probe spectra of Rb for $\nu_{LI}$ = 50, 250, and 1750 Hz, shown from top to bottom, respectively. (b) Example bi-Lorentzian fitting (blue line) on the pump-probe spectrum (black dots) at $\nu_{LI}$ = 1750 Hz. The spectra within the grey-shaded regions are used for noise level calculation. (c) The SNRs obtained with varying $\nu_{LI}$ from 10 to 1750 Hz for $T_{C}$ = 1000 ms (red triangles ) and 100 ms (blue circles). The dark current noise spectrum of the photodetector is shown as the grey-shaded trace.} 
    \label{fig:2}
\end{figure}

\begin{figure}[!ht]
    \centering
    \includegraphics[width=\columnwidth]{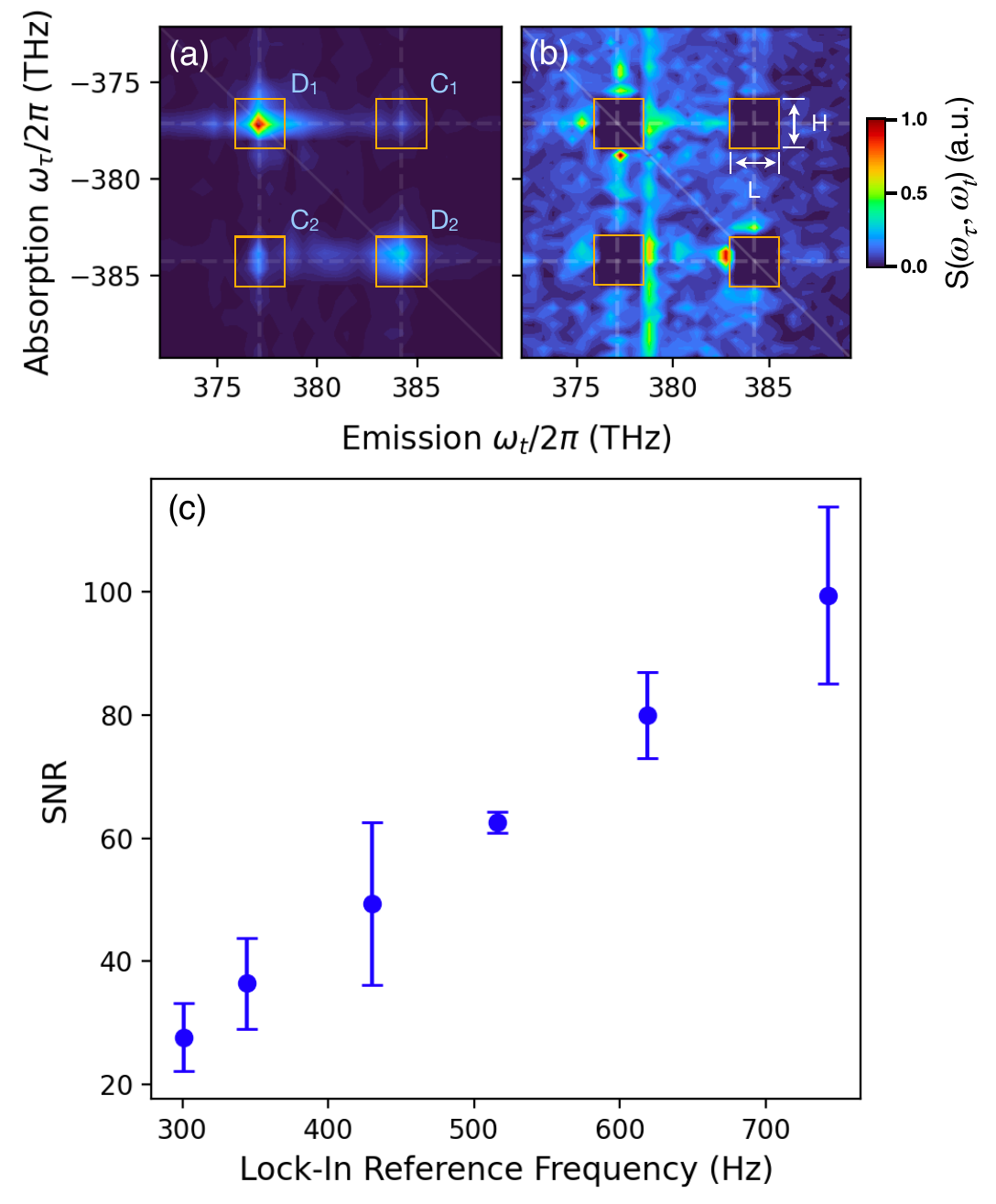}
    \caption{(a) Rephasing S1 2D spectrum for Rb obtained with $\nu_{LI} = 746$ Hz and $T_{C} = 1000$ ms. The rectangles illustrate the bounds of the signal areas. (b) The 2D spectrum with the signal in square boundaries removed for calculating the noise level. (c) The SNRs of 2D spectra obtained with different $\nu_{LI}$.} 
    \label{fig:3}
\end{figure}

The experimental schematic is shown in Fig. \ref{fig:1}(a). The laser source is a 5-kHz optical parametric amplifier (OPA) pumped by a regenerative amplifier, producing $<$ 70-fs laser pulses centered at 789 nm (380 THz) to excite both $D$ lines of Rb. The OPA output is sent into a collinear 2DCS spectrometer composed of two Mach-Zehnder interferometers nested within an outer Mach-Zehnder interferometer. Four fs pulses are generated and frequency-tagged with AOMs at modulation frequencies $\Omega_{i}$ ($i=A, B, C, D$). Interpulse time delays ($\tau$, $T$, and $t$) are introduced by three translational stages. The frequency-tagged pulses and their associated interpulse delays in a pulse sequence are shown as an inset of Fig. \ref{fig:1}(a). The four pulses are recombined in a co-propagating beam by three beamsplitters. The final beamsplitter produces two outputs of a four-pulse sequence, and each output has a total power of $\sim$2 mW. 

In our experiment, the fs pulses are used as both excitation pulses to generate nonlinear signals in a sample and a reference laser to provide an interferometric beating signal for lock-in detection. One output of a four-pulse sequence is used to excite the sample. 
The second output of the final beamsplitter acts as a reference frequency source for lock-in detection. This is done by measuring the interference signal of four pulses. Since the pulses can only interfere when they are overlapped in time, to have interference at a longer delay time, we integrated a pulse shaper in the optical setup to stretch the pulse duration to a few ps. 

The pulse shaper consists of an 1800 gr/mm grating, an adjustable mechanical slit, a 75-mm focal length cylindrical lens, a 25-mm focal length lens, and an amplified photodetector (PD$_{ref}$) to detect the reference. The fs laser pulse incident on the grating is dispersed into its component wavelengths and collimated with the cylindrical lens onto the mechanical slit. The slit output is focused with the 25-mm lens onto PD$_{ref}$. The slit selects a narrower spectrum and effectively increases the duration of the pulses. As shown in Fig. \ref{fig:1}(b), the spectrum of the OPA output is the black curve, which covers both $D_1$ and $D_2$ lines indicated by the red and blue dashed lines, respectively. The blue shaded area in Fig. \ref{fig:1}(b) is the spectrum measured after the slit with an opening of $\sim$250 $\mu$m. The fitted full-width half-maximum (FWHM) of the blue-shaded area is 1.34 THz (2.78 nm), which corresponds to a transform-limited pulse duration of 746 fs. If needed, the pulse can be further stretched with a smaller slit size. The output of the pulse shaper is used to generate the lock-in reference signal $\nu_{LI}$. 

We first performed the auto-correlation measurements of two pulses at a time while the other two pulses were blocked to accurately determine the stage positions for zero interpulse delays and measure the pulse duration. Since $\nu_{LI} < 0.5\nu_{rep}$, to retrieve the reference frequency, the PD$_{ref}$ output was passed through an analog 2-kHz low-pass filter (LPF) to remove the 5-kHz pulse train before further amplification. Figure \ref{fig:1}(c) shows the unfiltered PD$_{ref}$ output (blue line) with the 5-kHz pulse train and the filtered output (red line) demonstrating the slower modulated oscillation extracted for lock-in reference. A zoom-in of the pulse train is shown in Fig. \ref{fig:1}(d) prior to filtering. The 5-kHz pulse train observed here was removed by the LPF. After amplification, the filtered reference signal was sent into an audio mixer, where the desired beating frequency was digitally filtered with a band-pass filter (BPF), amplified, and then output to the lock-in amplifier (LIA) as a reference frequency. The demodulated LIA signal is output as the in-phase ($X$) and in-quadrature ($Y$) components of a complex signal $Z(t) = X(t) + iY(t)$ as a function of the interpulse time delay $t$. Fourier-transform of $Z(t)$ generates the frequency domain signal given by $S(\omega_{t}) = FFT[Z(t)]$. For 2DCS, the signal is a function of two time delays $t$ and $\tau$, making the frequency-domain signal $S(\omega_{t}, \omega_{\tau}) = FFT[Z(t, \tau)]$.

We use the auto-correlation measurement of pulses $A$ and $B$ as an example. Each AOM is modulated at a slightly different frequency around 80 MHz and the offset modulation we impose is given by $\nu_{i}$ ($i=A, B$). 
For convenience, the AOM modulation frequencies are denoted as $\Omega_{A,B}$ = $\nu_{0} + \nu_{A,B}$, where $\nu_{0}$ = 80 MHz and $\nu_{i}$ ($i=A,B$) is a small offset. The offsets $\nu_{i}$ are chosen so the beating frequency is below 2 kHz, the LPF cutoff frequency. Figure \ref{fig:1}(e) shows the auto-correlation spectra of pulses $A$ and $B$, acquired with $\nu_{A} = 700$ Hz and $\nu_{B} = 200$ Hz, thus the lock-in reference frequency $\nu_{LI} = 500$ Hz. The auto-correlation spectrum of pulses $A$ and $B$ is shown as black dots, and the pulse duration (FWHM) was determined to be $\Delta t = 69$ fs. After pulse-shaping, the auto-correlation is shown as blue dots, and the pulse duration became $\Delta t^{\prime} = 920$ fs. This measured duration is slightly longer than the transfer-limited duration (746 fs) due to chirp.  

We then performed two-pulse pump-probe experiments with pulses $A$ and $B$ while the remaining two were blocked. The sample studied is a Rb vapor cell heated at 160$^{\circ}$C. The pulses are incident on the window at 45$^{\circ}$. The reflected beam is blocked, and the emitted fluorescence is collected at the normal angle with a lens focusing onto a signal photodetector (PD$_{sig}$). The signal is then processed by a lock-in amplifier, in the same way as the auto-correlation measurement, to obtain a pump-probe spectrum S($\omega_{\tau}$). To study how the lock-in reference frequency $\nu_{LI}$ affects the SNR, pump-probe spectra were acquired with $\nu_{LI}$ varying between 10 and 1750 Hz. Two different integration times $T_{C}$ = 1000 and 100 ms were used in the lock-in amplifier. Examples of S($\omega_{\tau}$) for $\nu_{LI}$ = 50, 250, and 1750 Hz for $T_{C}$ = 1000 ms are shown in Fig. \ref{fig:2}(a). The spectra show two resonances corresponding to the two $D$-line transitions. 

The SNR is obtained for all pump-probe spectra. The example spectrum in Fig. \ref{fig:2}(b) taken at $\nu_{LI}$ = 1750 Hz illustrates how the SNR is calculated. The spectrum is fit with a bi-Lorentzian function to find the $D_{1}$ linewidth (FWHM) $\Gamma$ and peak amplitude $s_{1}$. The noise level is calculated from the grey-shaded portions (labeled $n_1$ and $n_2$) of the spectrum in Fig. \ref{fig:2}(b). The upper (lower) bound $\omega_{u}$ ($\omega_{l}$) for the $n_{1}$ ($n_2$) shaded region relative to the $D_{1}$ ($D_2$) center position $\omega_{1}$ ($\omega_{2}$) is determined as $\omega_{u} = \omega_{1} - 2\Gamma$ ($\omega_{l} = \omega_{2} + 2\Gamma$). Thus, all data points below $\omega_{u}$ and above $\omega_{l}$ were used to define the total noise floor. The noise level is defined as the standard deviation of the noise floor. The SNR is calculated as the ratio of the peak amplitude of the higher peak to the standard deviation of the noise floor.

The retrieved SNRs with various $\nu_{LI}$ values from 10 to 1750 Hz are shown in Fig. \ref{fig:2}(c) for $T_{C}$ = 1000 (red triangles) and 100 ms (blue circles). Error bars are calculated from the standard deviation of 3 consecutive pump-probe measurements. The grey trace is the dark current noise spectrum measured with a spectral analyzer from the output of PD$_{ref}$ with no incident light. For $\nu_{LI} >$ 250 Hz, the SNR is relatively constant with increasing $\nu_{LI}$ for $T_{C}=100$ ms, while showing a slight increase for $T_{C} = 1000$ ms. Lock-in reference frequencies $\nu_{LI} < 250$ Hz have a noticeable drop in SNR. At $\nu_{LI} = 10$ Hz, the SNR reaches the minimum values of 17 and 50 for $T_{C} = 100$ and 1000 ms, respectively. This result agrees well with the photodetector's dark current noise spectrum which shows a predominantly large increase below 250 Hz. 

To perform 2DCS experiments, the sample is now excited by 4 pulses modulated with AOM frequencies $\Omega_{A}$, $\Omega_{B}$, $\Omega_{C}$, and $\Omega_{D}$, respectively. The interpulse delay times are given the labels $\tau$, $T$, and $t$. The mixing time is set to $T = 140$ fs, while delays $\tau$ and $t$ are scanned to acquire a rephasing (S1) 2D spectrum. The setup and experiment are similar to the pump-probe experiment, except that pulses $C$ and $D$ are unblocked so four pulses are used. The lock-in reference frequency associated with an S1 2D spectrum is given by $\nu_{LI} = -\nu_{A} + \nu_{B} + \nu_{C} -\nu_{D}$. The initial modulations chosen were $\nu_{A}$ = 1.503 kHz, $\nu_{B}$ = 345 Hz, $\nu_{C}$ = 2.505 kHz, and $\nu_{D}$ = 604 Hz. This corresponds to a lock-in reference frequency of $\nu_{LI}$ = 743 Hz. An S1 2D spectrum at $\nu_{LI}$ = 743 Hz is shown in Fig. \ref{fig:3}(a). The spectrum has two diagonal peaks corresponding to Rb $D_{1}$ and $D_{2}$ resonances, and two cross peaks $C_{1}$ and $C_{2}$ due to the coupling of the two resonances.

To orchestrate the study of 2DCS SNR dependence on the lock-in reference frequency $\nu_{LI}$, the initial modulation frequencies $\nu_i$ mentioned above were multiplied by a factor $n$ to produce a set of new modulation frequencies, and thus a set of new lock-in reference frequencies $n\nu_{LI}$. This approach kept the relative ratio of each pair of modulations constant while systematically varying the lock-in reference frequency $\nu_{LI}$ for the dependence study. Similarly to the pump-probe experiment, beating frequencies between any two pules and the overall lock-in reference frequency ($\nu_{LI}$) for S1 2DCS were all kept below 2.5 kHz. Again, this was to avoid undersampling and due to the 2.0 kHz LPF. These limitations prevent $\nu_{LI}$ from going above 800 Hz. The S1 lock-in reference frequencies used were $\nu_{LI}$ = 301, 344, 430, 516, 619, and 743 Hz. The lock-in integration time $T_{C}$ for all measurements was kept at 1000 ms. A series of S1 2D spectra were acquired and analyzed for the SNR. 

The peak amplitude of the strongest peak (labelled $D_{1}$) in all S1 2D spectra was used to calculate the SNR. The noise level was analyzed by initially removing contributions of all signals from the four peaks in the 2D spectrum to produce a filtered 2D spectrum. This was achieved by setting boundaries centered on each resonance and setting all values within the bounds to zero, effectively leaving the noise in the spectrum. The boundary length ($L$) and height ($H$) are twice the 2D diagonal peak FWHM in the horizontal and vertical directions, respectively, as determined by Lorentzian fitting to the horizontal and vertical slices of the $D_{1}$ diagonal peak. The horizontal and vertical linewidths were found to be 1.1 THz and 0.94 THz from the 2D spectrum in Fig. \ref{fig:3}(a). The boundary values used for all 4 resonances were set to $(L,H) = (2.2, 1.88)$ THz. An example filtered spectrum with all signal contributions removed is shown in Fig. \ref{fig:3}(b). The spectrum is re-normalized only for plotting purposes and not for calculating the noise level. The noise level is calculated similarly to the approach used in the pump-probe experiment as the standard deviation of the noise floor. The SNR is the ratio of the $D_{1}$ peak amplitude to the standard deviation. The obtained SNRs for all $\nu_{LI}$ are plotted in Fig. \ref{fig:3}(c). Akin to the pump-probe case, error bars were determined from 3 consecutive measurements. The SNR increases continuously as $\nu_{LI}$ goes from 301 to 743 Hz. From this result, we determine the optimal lock-in reference frequency in our 2DCS experimental with a 5-KHz laser is around 743 Hz. The SNR is about 100 which is comparable to the optimal SNR in the pump-probe experiment. 

In conclusion, we successfully implemented collinear 2DCS based on fluorescence detection for a Rb vapor with a 5-kHz repetition rate fs laser. The fs pulses work as both excitation and reference for lock-in detection, with a grating-based pulse shaper to stretch the pulse duration to ps scale for reference signal. Both pump-probe and rephasing 2D spectra were obtained. The effectiveness of the technique was characterized by calculating the SNR for pump-probe and rephasing 2D spectra for different lock-in reference frequencies $\nu_{LI}$ and integration time $T_{C}$. In the case of pump-probe, the SNR decreases to the noise floor for $\nu_{LI} <$ 250 Hz, while a SNR plateau is reached for $\nu_{LI} >$ 250 Hz for $T_{C}$ = 100 and 1000 ms. The S1 2DCS SNR characterization shows a positive trend with increasing lock-in reference frequency. The successful demonstration of collinear 2DCS with a kHz laser paves the way for applying collinear 2DCS in scenarios (such as defect color centers in solids) that require a kHz repetition rate laser to utilize high pulse energy, wavelength tunability, low average power to avoid heating, and long waiting time between shots to allow a full relaxation back to the ground state.

We acknowledge the support by Army Research Office (W911NF-23-1-0195) and National Science Foundation (PHY-2216824).


\bibliography{apssamp}

\begin{thebibliography}{23}%
\makeatletter
\providecommand \@ifxundefined [1]{%
 \@ifx{#1\undefined}
}%
\providecommand \@ifnum [1]{%
 \ifnum #1\expandafter \@firstoftwo
 \else \expandafter \@secondoftwo
 \fi
}%
\providecommand \@ifx [1]{%
 \ifx #1\expandafter \@firstoftwo
 \else \expandafter \@secondoftwo
 \fi
}%
\providecommand \natexlab [1]{#1}%
\providecommand \enquote  [1]{``#1''}%
\providecommand \bibnamefont  [1]{#1}%
\providecommand \bibfnamefont [1]{#1}%
\providecommand \citenamefont [1]{#1}%
\providecommand \href@noop [0]{\@secondoftwo}%
\providecommand \href [0]{\begingroup \@sanitize@url \@href}%
\providecommand \@href[1]{\@@startlink{#1}\@@href}%
\providecommand \@@href[1]{\endgroup#1\@@endlink}%
\providecommand \@sanitize@url [0]{\catcode `\\12\catcode `\$12\catcode `\&12\catcode `\#12\catcode `\^12\catcode `\_12\catcode `\%12\relax}%
\providecommand \@@startlink[1]{}%
\providecommand \@@endlink[0]{}%
\providecommand \url  [0]{\begingroup\@sanitize@url \@url }%
\providecommand \@url [1]{\endgroup\@href {#1}{\urlprefix }}%
\providecommand \urlprefix  [0]{URL }%
\providecommand \Eprint [0]{\href }%
\providecommand \doibase [0]{https://doi.org/}%
\providecommand \selectlanguage [0]{\@gobble}%
\providecommand \bibinfo  [0]{\@secondoftwo}%
\providecommand \bibfield  [0]{\@secondoftwo}%
\providecommand \translation [1]{[#1]}%
\providecommand \BibitemOpen [0]{}%
\providecommand \bibitemStop [0]{}%
\providecommand \bibitemNoStop [0]{.\EOS\space}%
\providecommand \EOS [0]{\spacefactor3000\relax}%
\providecommand \BibitemShut  [1]{\csname bibitem#1\endcsname}%
\let\auto@bib@innerbib\@empty
\bibitem [{\citenamefont {Dai}\ \emph {et~al.}(2012)\citenamefont {Dai}, \citenamefont {Richter}, \citenamefont {Li}, \citenamefont {Bristow}, \citenamefont {Falvo}, \citenamefont {Mukamel},\ and\ \citenamefont {Cundiff}}]{Dai2012a}%
  \BibitemOpen
  \bibfield  {author} {\bibinfo {author} {\bibfnamefont {X.}~\bibnamefont {Dai}}, \bibinfo {author} {\bibfnamefont {M.}~\bibnamefont {Richter}}, \bibinfo {author} {\bibfnamefont {H.}~\bibnamefont {Li}}, \bibinfo {author} {\bibfnamefont {A.~D.}\ \bibnamefont {Bristow}}, \bibinfo {author} {\bibfnamefont {C.}~\bibnamefont {Falvo}}, \bibinfo {author} {\bibfnamefont {S.}~\bibnamefont {Mukamel}},\ and\ \bibinfo {author} {\bibfnamefont {S.~T.}\ \bibnamefont {Cundiff}},\ }\bibfield  {title} {\bibinfo {title} {{Two-Dimensional Double-Quantum Spectra Reveal Collective Resonances in an Atomic Vapor}},\ }\href {https://doi.org/10.1103/PhysRevLett.108.193201} {\bibfield  {journal} {\bibinfo  {journal} {Phys. Rev. Lett.}\ }\textbf {\bibinfo {volume} {108}},\ \bibinfo {pages} {193201} (\bibinfo {year} {2012})}\BibitemShut {NoStop}%
\bibitem [{\citenamefont {Gao}\ \emph {et~al.}(2016)\citenamefont {Gao}, \citenamefont {Cundiff},\ and\ \citenamefont {Li}}]{Gao:16}%
  \BibitemOpen
  \bibfield  {author} {\bibinfo {author} {\bibfnamefont {F.}~\bibnamefont {Gao}}, \bibinfo {author} {\bibfnamefont {S.~T.}\ \bibnamefont {Cundiff}},\ and\ \bibinfo {author} {\bibfnamefont {H.}~\bibnamefont {Li}},\ }\bibfield  {title} {\bibinfo {title} {{Probing dipole--dipole interaction in a rubidium gas via double-quantum 2D spectroscopy}},\ }\href {https://doi.org/10.1364/OL.41.002954} {\bibfield  {journal} {\bibinfo  {journal} {Opt. Lett.}\ }\textbf {\bibinfo {volume} {41}},\ \bibinfo {pages} {2954} (\bibinfo {year} {2016})}\BibitemShut {NoStop}%
\bibitem [{\citenamefont {Lomsadze}\ and\ \citenamefont {Cundiff}(2018)}]{PhysRevLett.120.233401}%
  \BibitemOpen
  \bibfield  {author} {\bibinfo {author} {\bibfnamefont {B.}~\bibnamefont {Lomsadze}}\ and\ \bibinfo {author} {\bibfnamefont {S.~T.}\ \bibnamefont {Cundiff}},\ }\bibfield  {title} {\bibinfo {title} {Frequency-comb based double-quantum two-dimensional spectrum identifies collective hyperfine resonances in atomic vapor induced by dipole-dipole interactions},\ }\href {https://doi.org/10.1103/PhysRevLett.120.233401} {\bibfield  {journal} {\bibinfo  {journal} {Phys. Rev. Lett.}\ }\textbf {\bibinfo {volume} {120}},\ \bibinfo {pages} {233401} (\bibinfo {year} {2018})}\BibitemShut {NoStop}%
\bibitem [{\citenamefont {Yu}\ \emph {et~al.}(2019{\natexlab{a}})\citenamefont {Yu}, \citenamefont {Titze}, \citenamefont {Zhu}, \citenamefont {Liu},\ and\ \citenamefont {Li}}]{Yu2018}%
  \BibitemOpen
  \bibfield  {author} {\bibinfo {author} {\bibfnamefont {S.}~\bibnamefont {Yu}}, \bibinfo {author} {\bibfnamefont {M.}~\bibnamefont {Titze}}, \bibinfo {author} {\bibfnamefont {Y.}~\bibnamefont {Zhu}}, \bibinfo {author} {\bibfnamefont {X.}~\bibnamefont {Liu}},\ and\ \bibinfo {author} {\bibfnamefont {H.}~\bibnamefont {Li}},\ }\bibfield  {title} {\bibinfo {title} {Long range dipole-dipole interaction in low-density atomic vapors probed by double-quantum two-dimensional coherent spectroscopy},\ }\href@noop {} {\bibfield  {journal} {\bibinfo  {journal} {Opt. Express}\ }\textbf {\bibinfo {volume} {27}},\ \bibinfo {pages} {28891} (\bibinfo {year} {2019}{\natexlab{a}})}\BibitemShut {NoStop}%
\bibitem [{\citenamefont {Yu}\ \emph {et~al.}(2019{\natexlab{b}})\citenamefont {Yu}, \citenamefont {Titze}, \citenamefont {Zhu}, \citenamefont {Liu},\ and\ \citenamefont {Li}}]{Yu2019}%
  \BibitemOpen
  \bibfield  {author} {\bibinfo {author} {\bibfnamefont {S.}~\bibnamefont {Yu}}, \bibinfo {author} {\bibfnamefont {M.}~\bibnamefont {Titze}}, \bibinfo {author} {\bibfnamefont {Y.}~\bibnamefont {Zhu}}, \bibinfo {author} {\bibfnamefont {X.}~\bibnamefont {Liu}},\ and\ \bibinfo {author} {\bibfnamefont {H.}~\bibnamefont {Li}},\ }\bibfield  {title} {\bibinfo {title} {{Observation of scalable and deterministic multi-atom Dicke states in an atomic vapor}},\ }\href {https://doi.org/10.1364/OL.44.002795} {\bibfield  {journal} {\bibinfo  {journal} {Opt. Lett.}\ }\textbf {\bibinfo {volume} {44}},\ \bibinfo {pages} {2795} (\bibinfo {year} {2019}{\natexlab{b}})}\BibitemShut {NoStop}%
\bibitem [{\citenamefont {Liang}\ and\ \citenamefont {Li}(2021)}]{Liang2021}%
  \BibitemOpen
  \bibfield  {author} {\bibinfo {author} {\bibfnamefont {D.}~\bibnamefont {Liang}}\ and\ \bibinfo {author} {\bibfnamefont {H.}~\bibnamefont {Li}},\ }\bibfield  {title} {\bibinfo {title} {Optical two-dimensional coherent spectroscopy of many-body dipole–dipole interactions and correlations in atomic vapors},\ }\href@noop {} {\bibfield  {journal} {\bibinfo  {journal} {J. Chem. Phys.}\ }\textbf {\bibinfo {volume} {154}},\ \bibinfo {pages} {214301} (\bibinfo {year} {2021})}\BibitemShut {NoStop}%
\bibitem [{\citenamefont {Yan}\ \emph {et~al.}(2022)\citenamefont {Yan}, \citenamefont {Revesz}, \citenamefont {Liang},\ and\ \citenamefont {Li}}]{Yan2022}%
  \BibitemOpen
  \bibfield  {author} {\bibinfo {author} {\bibfnamefont {J.}~\bibnamefont {Yan}}, \bibinfo {author} {\bibfnamefont {S.}~\bibnamefont {Revesz}}, \bibinfo {author} {\bibfnamefont {D.}~\bibnamefont {Liang}},\ and\ \bibinfo {author} {\bibfnamefont {H.}~\bibnamefont {Li}},\ }\bibfield  {title} {\bibinfo {title} {Broadband optical two-dimensional coherent spectroscopy of a rubidium atomic vapor},\ }\href@noop {} {\bibfield  {journal} {\bibinfo  {journal} {Phys. Rev. A}\ }\textbf {\bibinfo {volume} {105}},\ \bibinfo {pages} {052810} (\bibinfo {year} {2022})}\BibitemShut {NoStop}%
\bibitem [{\citenamefont {Liang}\ \emph {et~al.}(2022{\natexlab{a}})\citenamefont {Liang}, \citenamefont {Zhu},\ and\ \citenamefont {Li}}]{PhysRevLett.128.103601}%
  \BibitemOpen
  \bibfield  {author} {\bibinfo {author} {\bibfnamefont {D.}~\bibnamefont {Liang}}, \bibinfo {author} {\bibfnamefont {Y.}~\bibnamefont {Zhu}},\ and\ \bibinfo {author} {\bibfnamefont {H.}~\bibnamefont {Li}},\ }\bibfield  {title} {\bibinfo {title} {Collective resonance of $d$ states in rubidium atoms probed by optical two-dimensional coherent spectroscopy},\ }\href@noop {} {\bibfield  {journal} {\bibinfo  {journal} {Phys. Rev. Lett.}\ }\textbf {\bibinfo {volume} {128}},\ \bibinfo {pages} {103601} (\bibinfo {year} {2022}{\natexlab{a}})}\BibitemShut {NoStop}%
\bibitem [{\citenamefont {Landmesser}\ \emph {et~al.}(2023)\citenamefont {Landmesser}, \citenamefont {Sixt}, \citenamefont {Dulitz}, \citenamefont {Bruder},\ and\ \citenamefont {Stienkemeier}}]{Landmesser23}%
  \BibitemOpen
  \bibfield  {author} {\bibinfo {author} {\bibfnamefont {F.}~\bibnamefont {Landmesser}}, \bibinfo {author} {\bibfnamefont {T.}~\bibnamefont {Sixt}}, \bibinfo {author} {\bibfnamefont {K.}~\bibnamefont {Dulitz}}, \bibinfo {author} {\bibfnamefont {L.}~\bibnamefont {Bruder}},\ and\ \bibinfo {author} {\bibfnamefont {F.}~\bibnamefont {Stienkemeier}},\ }\bibfield  {title} {\bibinfo {title} {Two-dimensional electronic spectroscopy of an ultracold gas},\ }\href {https://doi.org/10.1364/OL.477301} {\bibfield  {journal} {\bibinfo  {journal} {Opt. Lett.}\ }\textbf {\bibinfo {volume} {48}},\ \bibinfo {pages} {473} (\bibinfo {year} {2023})}\BibitemShut {NoStop}%
\bibitem [{\citenamefont {Liang}\ \emph {et~al.}(2022{\natexlab{b}})\citenamefont {Liang}, \citenamefont {Rodriguez}, \citenamefont {Zhou}, \citenamefont {Zhu},\ and\ \citenamefont {Li}}]{Liang2022}%
  \BibitemOpen
  \bibfield  {author} {\bibinfo {author} {\bibfnamefont {D.}~\bibnamefont {Liang}}, \bibinfo {author} {\bibfnamefont {L.~S.}\ \bibnamefont {Rodriguez}}, \bibinfo {author} {\bibfnamefont {H.}~\bibnamefont {Zhou}}, \bibinfo {author} {\bibfnamefont {Y.}~\bibnamefont {Zhu}},\ and\ \bibinfo {author} {\bibfnamefont {H.}~\bibnamefont {Li}},\ }\bibfield  {title} {\bibinfo {title} {Optical two-dimensional coherent spectroscopy of cold atoms},\ }\href {https://doi.org/10.1364/OL.478793} {\bibfield  {journal} {\bibinfo  {journal} {Opt. Lett.}\ }\textbf {\bibinfo {volume} {47}},\ \bibinfo {pages} {6452} (\bibinfo {year} {2022}{\natexlab{b}})}\BibitemShut {NoStop}%
\bibitem [{\citenamefont {Moody}\ \emph {et~al.}(2015)\citenamefont {Moody}, \citenamefont {{Kavir Dass}}, \citenamefont {Hao}, \citenamefont {Chen}, \citenamefont {Li}, \citenamefont {Singh}, \citenamefont {Tran}, \citenamefont {Clark}, \citenamefont {Xu}, \citenamefont {Bergh{\"{a}}user}, \citenamefont {Malic}, \citenamefont {Knorr},\ and\ \citenamefont {Li}}]{Moody2015}%
  \BibitemOpen
  \bibfield  {author} {\bibinfo {author} {\bibfnamefont {G.}~\bibnamefont {Moody}}, \bibinfo {author} {\bibfnamefont {C.}~\bibnamefont {{Kavir Dass}}}, \bibinfo {author} {\bibfnamefont {K.}~\bibnamefont {Hao}}, \bibinfo {author} {\bibfnamefont {C.-H.}\ \bibnamefont {Chen}}, \bibinfo {author} {\bibfnamefont {L.-J.}\ \bibnamefont {Li}}, \bibinfo {author} {\bibfnamefont {A.}~\bibnamefont {Singh}}, \bibinfo {author} {\bibfnamefont {K.}~\bibnamefont {Tran}}, \bibinfo {author} {\bibfnamefont {G.}~\bibnamefont {Clark}}, \bibinfo {author} {\bibfnamefont {X.}~\bibnamefont {Xu}}, \bibinfo {author} {\bibfnamefont {G.}~\bibnamefont {Bergh{\"{a}}user}}, \bibinfo {author} {\bibfnamefont {E.}~\bibnamefont {Malic}}, \bibinfo {author} {\bibfnamefont {A.}~\bibnamefont {Knorr}},\ and\ \bibinfo {author} {\bibfnamefont {X.}~\bibnamefont {Li}},\ }\bibfield  {title} {\bibinfo {title} {{Intrinsic homogeneous linewidth and broadening mechanisms of excitons in monolayer transition metal dichalcogenides}},\ }\href
  {https://doi.org/10.1038/ncomms9315} {\bibfield  {journal} {\bibinfo  {journal} {Nat. Commun.}\ }\textbf {\bibinfo {volume} {6}},\ \bibinfo {pages} {8315} (\bibinfo {year} {2015})}\BibitemShut {NoStop}%
\bibitem [{\citenamefont {Titze}\ \emph {et~al.}(2018)\citenamefont {Titze}, \citenamefont {Li}, \citenamefont {Zhang}, \citenamefont {Ajayan},\ and\ \citenamefont {Li}}]{Titze2018}%
  \BibitemOpen
  \bibfield  {author} {\bibinfo {author} {\bibfnamefont {M.}~\bibnamefont {Titze}}, \bibinfo {author} {\bibfnamefont {B.}~\bibnamefont {Li}}, \bibinfo {author} {\bibfnamefont {X.}~\bibnamefont {Zhang}}, \bibinfo {author} {\bibfnamefont {P.~M.}\ \bibnamefont {Ajayan}},\ and\ \bibinfo {author} {\bibfnamefont {H.}~\bibnamefont {Li}},\ }\bibfield  {title} {\bibinfo {title} {{Intrinsic coherence time of trions in monolayer MoSe2 measured via two-dimensional coherent spectroscopy}},\ }\href {https://doi.org/10.1103/PhysRevMaterials.2.054001} {\bibfield  {journal} {\bibinfo  {journal} {Phys. Rev. Materials}\ }\textbf {\bibinfo {volume} {2}},\ \bibinfo {pages} {054001} (\bibinfo {year} {2018})}\BibitemShut {NoStop}%
\bibitem [{\citenamefont {Huang}\ \emph {et~al.}(2023)\citenamefont {Huang}, \citenamefont {Sampson}, \citenamefont {Ni}, \citenamefont {Liu}, \citenamefont {Liang}, \citenamefont {Watanabe}, \citenamefont {Taniguchi}, \citenamefont {Li}, \citenamefont {Martin}, \citenamefont {Levinsen}, \citenamefont {Parish}, \citenamefont {Tutuc}, \citenamefont {Efimkin},\ and\ \citenamefont {Li}}]{PhysRevX.13.011029}%
  \BibitemOpen
  \bibfield  {author} {\bibinfo {author} {\bibfnamefont {D.}~\bibnamefont {Huang}}, \bibinfo {author} {\bibfnamefont {K.}~\bibnamefont {Sampson}}, \bibinfo {author} {\bibfnamefont {Y.}~\bibnamefont {Ni}}, \bibinfo {author} {\bibfnamefont {Z.}~\bibnamefont {Liu}}, \bibinfo {author} {\bibfnamefont {D.}~\bibnamefont {Liang}}, \bibinfo {author} {\bibfnamefont {K.}~\bibnamefont {Watanabe}}, \bibinfo {author} {\bibfnamefont {T.}~\bibnamefont {Taniguchi}}, \bibinfo {author} {\bibfnamefont {H.}~\bibnamefont {Li}}, \bibinfo {author} {\bibfnamefont {E.}~\bibnamefont {Martin}}, \bibinfo {author} {\bibfnamefont {J.}~\bibnamefont {Levinsen}}, \bibinfo {author} {\bibfnamefont {M.~M.}\ \bibnamefont {Parish}}, \bibinfo {author} {\bibfnamefont {E.}~\bibnamefont {Tutuc}}, \bibinfo {author} {\bibfnamefont {D.~K.}\ \bibnamefont {Efimkin}},\ and\ \bibinfo {author} {\bibfnamefont {X.}~\bibnamefont {Li}},\ }\bibfield  {title} {\bibinfo {title} {Quantum dynamics of attractive and repulsive polarons in a doped ${\mathrm{mose}}_{2}$
  monolayer},\ }\href {https://doi.org/10.1103/PhysRevX.13.011029} {\bibfield  {journal} {\bibinfo  {journal} {Phys. Rev. X}\ }\textbf {\bibinfo {volume} {13}},\ \bibinfo {pages} {011029} (\bibinfo {year} {2023})}\BibitemShut {NoStop}%
\bibitem [{\citenamefont {Smallwood}\ \emph {et~al.}(2021)\citenamefont {Smallwood}, \citenamefont {Ulbricht}, \citenamefont {Day}, \citenamefont {Schröder}, \citenamefont {Bates}, \citenamefont {Autry}, \citenamefont {Diederich}, \citenamefont {Bielejec}, \citenamefont {Siemens},\ and\ \citenamefont {Cundiff}}]{Smallwood2021}%
  \BibitemOpen
  \bibfield  {author} {\bibinfo {author} {\bibfnamefont {C.~L.}\ \bibnamefont {Smallwood}}, \bibinfo {author} {\bibfnamefont {R.}~\bibnamefont {Ulbricht}}, \bibinfo {author} {\bibfnamefont {M.~W.}\ \bibnamefont {Day}}, \bibinfo {author} {\bibfnamefont {T.}~\bibnamefont {Schröder}}, \bibinfo {author} {\bibfnamefont {K.~M.}\ \bibnamefont {Bates}}, \bibinfo {author} {\bibfnamefont {T.~M.}\ \bibnamefont {Autry}}, \bibinfo {author} {\bibfnamefont {G.}~\bibnamefont {Diederich}}, \bibinfo {author} {\bibfnamefont {E.}~\bibnamefont {Bielejec}}, \bibinfo {author} {\bibfnamefont {M.~E.}\ \bibnamefont {Siemens}},\ and\ \bibinfo {author} {\bibfnamefont {S.~T.}\ \bibnamefont {Cundiff}},\ }\bibfield  {title} {\bibinfo {title} {{Hidden Silicon-Vacancy Centers in Diamond}},\ }\href@noop {} {\bibfield  {journal} {\bibinfo  {journal} {Phys. Rev. Lett.}\ }\textbf {\bibinfo {volume} {126}},\ \bibinfo {pages} {213601} (\bibinfo {year} {2021})}\BibitemShut {NoStop}%
\bibitem [{\citenamefont {Li}\ \emph {et~al.}(2023)\citenamefont {Li}, \citenamefont {Lomsadze}, \citenamefont {Moody}, \citenamefont {Smallwood},\ and\ \citenamefont {Cundiff}}]{MDCSBook2023}%
  \BibitemOpen
  \bibfield  {author} {\bibinfo {author} {\bibfnamefont {H.}~\bibnamefont {Li}}, \bibinfo {author} {\bibfnamefont {B.}~\bibnamefont {Lomsadze}}, \bibinfo {author} {\bibfnamefont {G.}~\bibnamefont {Moody}}, \bibinfo {author} {\bibfnamefont {C.}~\bibnamefont {Smallwood}},\ and\ \bibinfo {author} {\bibfnamefont {S.}~\bibnamefont {Cundiff}},\ }\href {https://doi.org/10.1093/oso/9780192843869.001.0001} {\emph {\bibinfo {title} {{Optical Multidimensional Coherent Spectroscopy}}}}\ (\bibinfo  {publisher} {Oxford University Press},\ \bibinfo {year} {2023})\ \Eprint {https://arxiv.org/abs/https://academic.oup.com/book/45704/book-pdf/50322612/9780192657626\_web.pdf} {https://academic.oup.com/book/45704/book-pdf/50322612/9780192657626\_web.pdf} \BibitemShut {NoStop}%
\bibitem [{\citenamefont {Brixner}\ \emph {et~al.}(2004)\citenamefont {Brixner}, \citenamefont {Stiopkin},\ and\ \citenamefont {Fleming}}]{Brixner:04}%
  \BibitemOpen
  \bibfield  {author} {\bibinfo {author} {\bibfnamefont {T.}~\bibnamefont {Brixner}}, \bibinfo {author} {\bibfnamefont {I.~V.}\ \bibnamefont {Stiopkin}},\ and\ \bibinfo {author} {\bibfnamefont {G.~R.}\ \bibnamefont {Fleming}},\ }\bibfield  {title} {\bibinfo {title} {Tunable two-dimensional femtosecond spectroscopy},\ }\href {https://doi.org/10.1364/OL.29.000884} {\bibfield  {journal} {\bibinfo  {journal} {Opt. Lett.}\ }\textbf {\bibinfo {volume} {29}},\ \bibinfo {pages} {884} (\bibinfo {year} {2004})}\BibitemShut {NoStop}%
\bibitem [{\citenamefont {Cowan}\ \emph {et~al.}(2004)\citenamefont {Cowan}, \citenamefont {Ogilvie},\ and\ \citenamefont {Miller}}]{COWAN2004184}%
  \BibitemOpen
  \bibfield  {author} {\bibinfo {author} {\bibfnamefont {M.}~\bibnamefont {Cowan}}, \bibinfo {author} {\bibfnamefont {J.}~\bibnamefont {Ogilvie}},\ and\ \bibinfo {author} {\bibfnamefont {R.}~\bibnamefont {Miller}},\ }\bibfield  {title} {\bibinfo {title} {Two-dimensional spectroscopy using diffractive optics based phased-locked photon echoes},\ }\href {https://doi.org/https://doi.org/10.1016/j.cplett.2004.01.027} {\bibfield  {journal} {\bibinfo  {journal} {Chemical Physics Letters}\ }\textbf {\bibinfo {volume} {386}},\ \bibinfo {pages} {184} (\bibinfo {year} {2004})}\BibitemShut {NoStop}%
\bibitem [{\citenamefont {Bristow}\ \emph {et~al.}(2009)\citenamefont {Bristow}, \citenamefont {Karaiskaj}, \citenamefont {Dai}, \citenamefont {Zhang}, \citenamefont {Carlsson}, \citenamefont {Hagen}, \citenamefont {Jimenez},\ and\ \citenamefont {Cundiff}}]{Bristow2009}%
  \BibitemOpen
  \bibfield  {author} {\bibinfo {author} {\bibfnamefont {A.~D.}\ \bibnamefont {Bristow}}, \bibinfo {author} {\bibfnamefont {D.}~\bibnamefont {Karaiskaj}}, \bibinfo {author} {\bibfnamefont {X.}~\bibnamefont {Dai}}, \bibinfo {author} {\bibfnamefont {T.}~\bibnamefont {Zhang}}, \bibinfo {author} {\bibfnamefont {C.}~\bibnamefont {Carlsson}}, \bibinfo {author} {\bibfnamefont {K.~R.}\ \bibnamefont {Hagen}}, \bibinfo {author} {\bibfnamefont {R.}~\bibnamefont {Jimenez}},\ and\ \bibinfo {author} {\bibfnamefont {S.~T.}\ \bibnamefont {Cundiff}},\ }\bibfield  {title} {\bibinfo {title} {{A versatile ultrastable platform for optical multidimensional Fourier-transform spectroscopy}},\ }\href {https://doi.org/10.1063/1.3184103} {\bibfield  {journal} {\bibinfo  {journal} {Review of Scientific Instruments}\ }\textbf {\bibinfo {volume} {80}},\ \bibinfo {pages} {073108} (\bibinfo {year} {2009})},\ \Eprint {https://arxiv.org/abs/https://pubs.aip.org/aip/rsi/article-pdf/doi/10.1063/1.3184103/9927374/073108\_1\_online.pdf}
  {https://pubs.aip.org/aip/rsi/article-pdf/doi/10.1063/1.3184103/9927374/073108\_1\_online.pdf} \BibitemShut {NoStop}%
\bibitem [{\citenamefont {Turner}\ \emph {et~al.}(2011)\citenamefont {Turner}, \citenamefont {Stone}, \citenamefont {Gundogdu},\ and\ \citenamefont {Nelson}}]{Turner2011}%
  \BibitemOpen
  \bibfield  {author} {\bibinfo {author} {\bibfnamefont {D.~B.}\ \bibnamefont {Turner}}, \bibinfo {author} {\bibfnamefont {K.~W.}\ \bibnamefont {Stone}}, \bibinfo {author} {\bibfnamefont {K.}~\bibnamefont {Gundogdu}},\ and\ \bibinfo {author} {\bibfnamefont {K.~A.}\ \bibnamefont {Nelson}},\ }\bibfield  {title} {\bibinfo {title} {{Invited Article: The coherent optical laser beam recombination technique (COLBERT) spectrometer: Coherent multidimensional spectroscopy made easier}},\ }\href {https://doi.org/10.1063/1.3624752} {\bibfield  {journal} {\bibinfo  {journal} {Review of Scientific Instruments}\ }\textbf {\bibinfo {volume} {82}},\ \bibinfo {pages} {081301} (\bibinfo {year} {2011})},\ \Eprint {https://arxiv.org/abs/https://pubs.aip.org/aip/rsi/article-pdf/doi/10.1063/1.3624752/15896159/081301\_1\_online.pdf} {https://pubs.aip.org/aip/rsi/article-pdf/doi/10.1063/1.3624752/15896159/081301\_1\_online.pdf} \BibitemShut {NoStop}%
\bibitem [{\citenamefont {Tekavec}\ \emph {et~al.}(2007)\citenamefont {Tekavec}, \citenamefont {Lott},\ and\ \citenamefont {Marcus}}]{Tekavec2007}%
  \BibitemOpen
  \bibfield  {author} {\bibinfo {author} {\bibfnamefont {P.~F.}\ \bibnamefont {Tekavec}}, \bibinfo {author} {\bibfnamefont {G.~A.}\ \bibnamefont {Lott}},\ and\ \bibinfo {author} {\bibfnamefont {A.~H.}\ \bibnamefont {Marcus}},\ }\bibfield  {title} {\bibinfo {title} {{Fluorescence-detected two-dimensional electronic coherence spectroscopy by acousto-optic phase modulation}},\ }\href {https://doi.org/10.1063/1.2800560} {\bibfield  {journal} {\bibinfo  {journal} {The Journal of Chemical Physics}\ }\textbf {\bibinfo {volume} {127}},\ \bibinfo {pages} {214307} (\bibinfo {year} {2007})},\ \Eprint {https://arxiv.org/abs/https://pubs.aip.org/aip/jcp/article-pdf/doi/10.1063/1.2800560/15407120/214307\_1\_online.pdf} {https://pubs.aip.org/aip/jcp/article-pdf/doi/10.1063/1.2800560/15407120/214307\_1\_online.pdf} \BibitemShut {NoStop}%
\bibitem [{\citenamefont {Nardin}\ \emph {et~al.}(2013)\citenamefont {Nardin}, \citenamefont {Autry}, \citenamefont {Silverman},\ and\ \citenamefont {Cundiff}}]{Nardin2013}%
  \BibitemOpen
  \bibfield  {author} {\bibinfo {author} {\bibfnamefont {G.}~\bibnamefont {Nardin}}, \bibinfo {author} {\bibfnamefont {T.~M.}\ \bibnamefont {Autry}}, \bibinfo {author} {\bibfnamefont {K.~L.}\ \bibnamefont {Silverman}},\ and\ \bibinfo {author} {\bibfnamefont {S.~T.}\ \bibnamefont {Cundiff}},\ }\bibfield  {title} {\bibinfo {title} {Multidimensional coherent photocurrent spectroscopy of a semiconductor nanostructure},\ }\href {https://doi.org/10.1364/OE.21.028617} {\bibfield  {journal} {\bibinfo  {journal} {Opt. Express}\ }\textbf {\bibinfo {volume} {21}},\ \bibinfo {pages} {28617} (\bibinfo {year} {2013})}\BibitemShut {NoStop}%
\bibitem [{\citenamefont {Wagner}\ \emph {et~al.}(2005)\citenamefont {Wagner}, \citenamefont {Li}, \citenamefont {Semmlow},\ and\ \citenamefont {Warren}}]{Wagner:05}%
  \BibitemOpen
  \bibfield  {author} {\bibinfo {author} {\bibfnamefont {W.}~\bibnamefont {Wagner}}, \bibinfo {author} {\bibfnamefont {C.}~\bibnamefont {Li}}, \bibinfo {author} {\bibfnamefont {J.}~\bibnamefont {Semmlow}},\ and\ \bibinfo {author} {\bibfnamefont {W.~S.}\ \bibnamefont {Warren}},\ }\bibfield  {title} {\bibinfo {title} {Rapid phase-cycled two-dimensional optical spectroscopy in fluorescence and transmission mode},\ }\href {https://doi.org/10.1364/OPEX.13.003697} {\bibfield  {journal} {\bibinfo  {journal} {Opt. Express}\ }\textbf {\bibinfo {volume} {13}},\ \bibinfo {pages} {3697} (\bibinfo {year} {2005})}\BibitemShut {NoStop}%
\bibitem [{\citenamefont {Bruder}\ \emph {et~al.}(2018)\citenamefont {Bruder}, \citenamefont {Binz},\ and\ \citenamefont {Stienkemeier}}]{Bruder:18}%
  \BibitemOpen
  \bibfield  {author} {\bibinfo {author} {\bibfnamefont {L.}~\bibnamefont {Bruder}}, \bibinfo {author} {\bibfnamefont {M.}~\bibnamefont {Binz}},\ and\ \bibinfo {author} {\bibfnamefont {F.}~\bibnamefont {Stienkemeier}},\ }\bibfield  {title} {\bibinfo {title} {Phase-synchronous undersampling in nonlinear spectroscopy},\ }\href {https://doi.org/10.1364/OL.43.000875} {\bibfield  {journal} {\bibinfo  {journal} {Opt. Lett.}\ }\textbf {\bibinfo {volume} {43}},\ \bibinfo {pages} {875} (\bibinfo {year} {2018})}\BibitemShut {NoStop}%
\end{thebibliography}%

\end{document}